\documentclass[jcp,aip,reprint]{revtex4-1}

\usepackage{amsmath}
\usepackage{amsfonts}
\usepackage{amssymb}
\usepackage{graphicx}

\usepackage{subfigure}

\usepackage{algorithm}
\usepackage{algorithmicx}
\usepackage{algpseudocode}

\usepackage{bbm}

\newcommand{\nrm}[1]{\|#1\|}
\newcommand{\vc}[1]{\mathbf{#1}}
\newcommand{\leaf}{{\cal L}}

\begin{document}
\author{Markus Kowalewski}
\email{mkowalew@uci.edu}
\affiliation{Department of Information Technology, Uppsala University, Box~337, SE-751~05 Uppsala, Sweden}
\affiliation{Now at: Chemistry Department, University of California, Irvine, California 92697-2025, United States}

\author{Elisabeth Larsson}
\affiliation{Department of Information Technology, Uppsala University, Box~337, SE-751~05 Uppsala, Sweden}

\author{Alfa Heryudono}
\affiliation{Department of Mathematics, University of Massachusetts Dartmouth,
Dartmouth, Massachusetts 02747, United States}

\title{An adaptive interpolation scheme for molecular potential energy surfaces}
\begin{abstract}
The calculation of potential energy surfaces for quantum dynamics can be a time consuming task -- especially when a high level of theory for
the electronic structure calculation is required.
We propose an adaptive interpolation algorithm based on polyharmonic splines (PHS) combined with a partition of unity (PU)
approach. The adaptive node refinement allows to greatly reduce
the number of sample points by employing a local error estimate.
The algorithm and its scaling behavior is evaluated for a model
function in 2, 3  and 4 dimensions.
The developed algorithm allows for a more rapid and reliable interpolation of a potential energy surface within a given accuracy compared to the non-adaptive version.  
\end{abstract}

\maketitle

\section{Introduction}
Molecular potential energy surfaces (PES) are frequently
used in theoretical chemistry to solve the Schr\"odinger equation
for the nuclei.
The task of calculating those PESs is routinely carried out by 
quantum chemistry (QC) programs for e.g., the calculation of infrared spectra.
The PES is represented in the simplest possible way by harmonic oscillators through the second derivatives of the energy with respect to the nuclear coordinates.
While this method is straightforward, for any method that goes beyond
a harmonic approximation there are some different options.
High resolution spectroscopy and chemical reaction dynamics rely on a global
representation of the PES, often realized using analytical
functions \cite{Quack}. While analytical functions are an efficient
way to describe molecular systems around an equilibrium geometry,
reaction dynamics explores configurations far off the equilibrium structures. 

The solution of the time dependent Schr\"odinger
equation (TDSE) for reaction dynamics usually relies on the Born-Oppenheimer
approximation \cite{Tannor}, meaning that the PES for solving the
the TDSE in the space of the nuclear degrees of freedom is obtained
from quantum chemistry methods. The shape of the potential energy landscape
can become fairly complex, which can make the use of analytical functions
quite difficult.
In principle one could calculate the needed grid by directly calculating
every point with an ab initio method. However, this is usually too expensive
since a quantum dynamics simulation based on Fourier or finite difference
methods requires a fairly dense grid of sample points \cite{Tannor}.  

The two main categories for the representation of PESs can be divided into
fitting methods and interpolation methods. Fitting methods include simple polynomial representations, many-body polynomials  \cite{Varandas07apc}
and a broad variety of neural networks \cite{Manzhos15ijqc,Behler15ijqc}.
Interpolation methods provide the possibility to represent
the PES without the necessity of prior knowledge of its shape, and reproduce
the function exactly at the sample points.
Interpolation methods based on 
Shepard's method \cite{Shepard68,Tishchenko10jcp,Moyano04} perform
well, when combined with energy gradients and second derivatives.
Interpolating moving least squares \cite{Majumder16mp,Ishida99cpl}
target efficient and automated calculation of PESs.
Other interpolation methods which have been applied for PESs include
cubic splines \cite{McLaughlin73jcp,Xu05jcp},
Hilbert space reproducing kernels \cite{Hollebeek01jcp}, which
are a radial basis function (RBF) variant, or
are specially designed for inter-surface crossings \cite{Opalka13}.
A very general approach for two-dimensional surfaces is the
thin plate spline interpolation \cite{Fasshauer} which has been successfully
applied to the quantum dynamics of reactive systems \cite{Geppert03jcp,Kowalewski14JPCA}.
However, already in three dimensions the number of sample points can become large enough to make the solution of the underlying linear system of equations computationally overly expensive.  

In this paper we present an interpolation approach which
is driven by the practical need for an efficient interpolation scheme
within a few dimensions in combination with high level QC methods
(e.g. MPn, coupled cluster, configuration interaction \cite{Szabo}).
The main objectives motivating the development are:
The minimization of the number of sample points necessary to obtain an interpolant of given accuracy.
This reduces the number of ab initio calculations, since those can take up a major amount of the calculation time.
The scheme should should avoid the ${\cal O} (N^3)$ scaling behavior of linear solvers and thus be able to handle large numbers of sample points efficiently.
The scheme should yield reliable PESs by providing error control without
prior knowledge of the shape of the PES.
Moreover, it should not rely on the use of energy gradients or second derivatives,
which might not available or too expensive for some high-level QC methods.
The approach we use is based on polyharmonic splines (PHSs) \cite{Fasshauer} in combination with a PU approach \cite{Wendland02} to allow for a larger number of sample points. The PHSs can be set up such that they provide smooth surfaces
by avoiding oscillatory behavior in between sample points.
An iterative refinement scheme reduces the number
of sample points and allows for setting an error tolerance.

The aim of this paper is to provide a scalable method for a rapid and
reliable interpolation of low dimensional PESs solely based on single point energy calculation without the necessity of prior knowledge of its qualitative features.

\section{Method}
In the following the building blocks for the algorithm will be discussed.
In sec. \ref{sec:interpolation} we briefly review the PHS interpolation and
its combination with a PU scheme. In sec.~\ref{sec:refinement}
the algorithm for the local refinement scheme is introduced and then followed
by an experimental analysis of the algorithm's performance in sec.~\ref{sec:results}.

\subsection{The interpolation scheme}
\label{sec:interpolation}
The basic interpolation method has to be chosen such that it fits the properties
of the function it should approximate.
In the absence of avoided crossings and conical intersections
PESs are smooth functions with continuous derivatives. This paper will focus
on the case where the potential is represented by a smooth function. Cusps and
other features involving discontinues in the higher derivatives need special
attention which is beyond the scope of this paper.
Thin plate splines \cite{TPS0,TPS} have been used successfully in the past to reliably interpolate two-dimensional PESs from a coarse grid obtained by
QC calculations to much finer grids as needed for TDSE calculations.
thin-plate splines were originally designed to minimize the bending energy of a sheet of metal
forced out of plane configuration at a certain number of points. 
Mathematically this is achieved by minimizing the norm over the second derivatives of the interpolant. 
This  avoids oscillatory behavior in between the sample points, which is a basic requirement for a robust interpolation scheme suitable for PESs.
The generalization of thin-plate splines to higher dimensions and to
higher orders are the poly harmonic splines (PHS) \cite{Fasshauer}.
The properties, of minimizing the bending energies of second order and higher, is preserved
thus avoiding oscillatory behavior also in higher dimensions than two.
The general PHS interpolant $u(\mathbf{x})$ is given by:
\begin{equation}
\label{eq:PH}
 u(\mathbf{x}) = \sum_{j=1}^{N} \lambda_j\phi_j^m(\vc{x}) + {\cal P}^m(\vc{x})\,,
\end{equation}
where $u$ is defined by a set $X = \{\vc{x}_{c,1},\ldots,\vc{x}_{c,N}\}$ of
$N$ sample points and $\vc{x}=(x_1,\ldots, x_d)$ is a
vector in $\mathbb{R}^d$.
The basis functions of a PHS are conditionally positive definite RBFs $\phi_j^m(\vc{x})$ of order $m$ of the form:
\begin{equation}
\phi_j^m(\vc{x})\equiv\phi^m(\nrm{\vc{x} - \vc{x}_j}_2) = 
\begin{cases}
r^{2m-2}\ln(r) &d \mbox{ even},\\
r^{2m-1} &d \mbox{ odd}.\\
\end{cases}
\end{equation}
Here $r$ is defined by the euclidean distance $\|\cdot\|_2$.
The PHS also includes a multivariate polynomial term
\begin{equation}
\label{eq:P}
{\cal P}^m(\vc{x}) = \sum_{|p|=0}^{m-1} \alpha_p \vc{x}^p\,,
\end{equation}
where $p=(p_1,\ldots,p_d)$ is a multi-index of positive integers $\mathbb{N}_0^+$ such that $|p|=p_1+\cdots+p_d$ and $\alpha_p$ are the polynomial coefficients.
As an example, for a polynomial with order $m=3$ and dimension $d=2$ this corresponds
to the multivariate second order polynomial
\begin{equation}
{\cal P}^3(\vc{x}) = \alpha_1 + \alpha_2 x_1 + \alpha_3 x_2 + \alpha_4 x_1^2 
+ \alpha_5 x_1 x_2 + \alpha_6 x_2^2 \,.
\end{equation}
The coefficients $\lambda$ and $\alpha$ defining the interpolant from eq.~\eqref{eq:PH} and~\eqref{eq:P} are obtained by solving the linear system
\begin{equation}
\begin{pmatrix}
A& P\\
P^T& 0
\end{pmatrix}
\begin{pmatrix}
\lambda\\
\alpha
\end{pmatrix} = 
\begin{pmatrix}
f\\
0
\end{pmatrix}\,.
\label{eq:sys}
\end{equation}
The matrix elements of $A^{N\times N}$ represent the RBFs and are defined as $A_{ij}=\phi^m(\nrm{\vc{x}_i - \vc{x}_j}_2)$.
The vector ${f}=(f_1,\ldots,f_N)$ contains the function values $f(\vc{x}_j)$ at the sample points $\vc{x}_j$.
The polynomimal part of \eqref{eq:PH} is represented by the block
$P^{N\times s}$:
\begin{equation}
P = 
\begin{pmatrix}
1& \vc{x}_1^{p_1}& \ldots & \vc{x}_1^{p_{s-1}}\\
\vdots & \vdots &       & \vdots\,\,\,\\
1 & \vc{x}_N^{p_1} & \ldots & \vc{x}_N^{p_{s-1}}\\
\end{pmatrix}\,.
\end{equation}
The number of columns in $P$ are given by the binomial coefficient $s=\binom{m-1+d}{d}$. Thus a linear problem of size $N+s$ has to be solved. Due to the polynomial part 
of eq.~\eqref{eq:PH} a polynomial of degree $m-1$ can be represented exact by the
interpolant.

The PHS interpolant from eq.~\eqref{eq:PH} can  already be used
for the approximation of a PES with a moderate number of sample points.
However, the number of sample points that
can be used in a practical scenario is limited by its scaling behavior.
For dense coefficient matrices, linear solvers
usually scale with ${\cal O}(N^3)$ meaning that with a large number of sample points, the computational time can quickly become too large.
Also, for a large number $N$ the condition number of $A$ increases and the linear system becomes ill conditioned, thus posing numerical problems.
A solution to this problem is to break up the global nature of eq.~\eqref{eq:PH}
by splitting up the interpolation domain in sub-domains of smaller sizes
which can then be handled independently in the PHS interpolation.

\begin{figure}
\includegraphics[width=8.5cm]{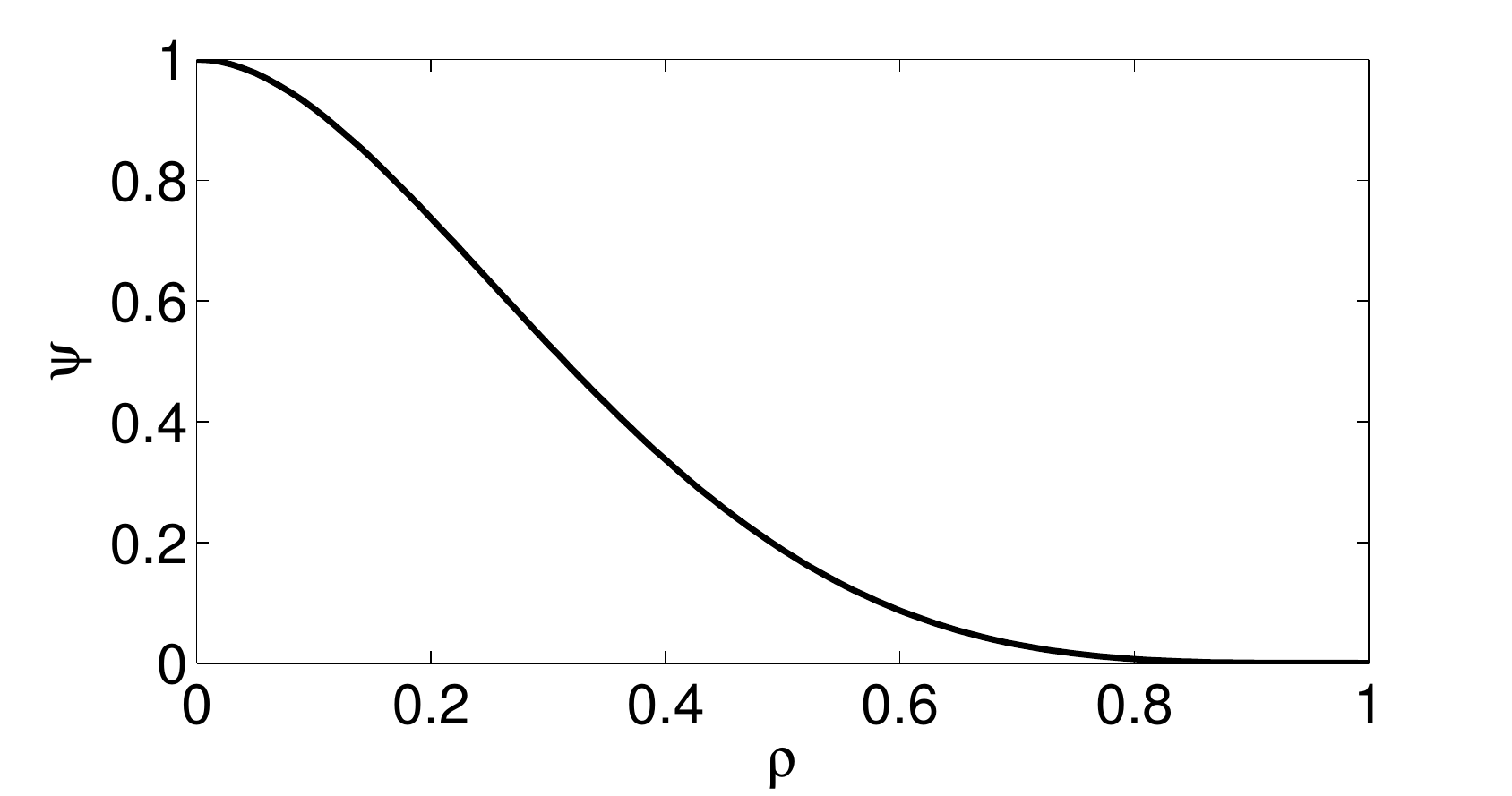}
\caption{Illustration of Wendland's $C^2$ function from eq.~\eqref{eq:psi} for two and three dimensions.}
\label{fig:WC2}
\end{figure}

\begin{figure}
\includegraphics[width=8.5cm]{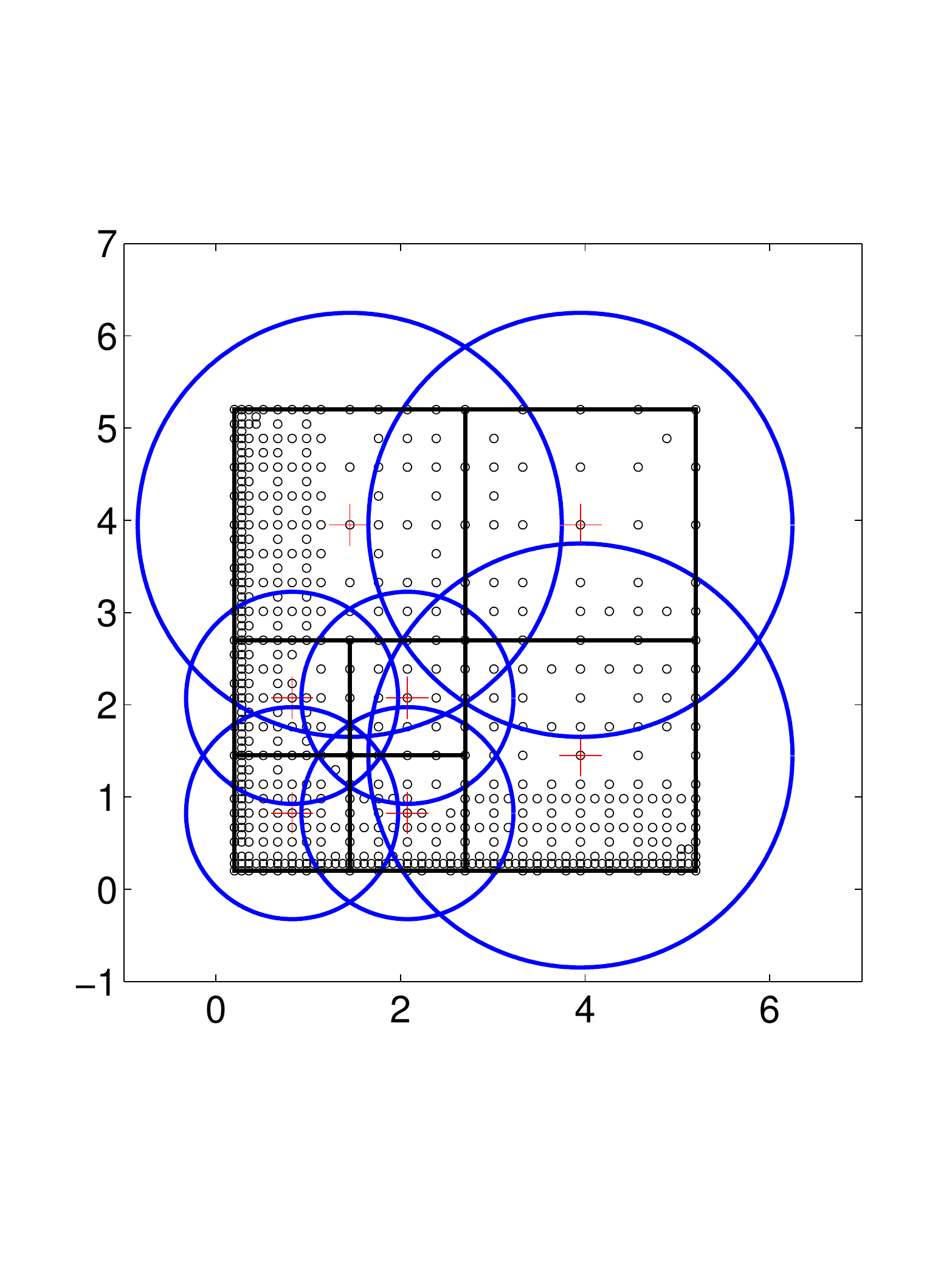}
\caption{Covering of a domain by circular patches~\cite{Her07,Saf15}. The domain is first subdivided into boxes (black lines) such that the number of data points (black circles) in each box does not exceed a given limit.
The patches are then formed as circles that that enclose each box
(box and circle centers, red crosses).
The circles (or hyperspheres in the general case) are enlarged by a fixed factor
to ensure a sufficient overlap between the patches (blue circles).}
\label{fig:part}
\end{figure}

There are several different methods available allowing for localization.
Compactly supported RBFs can be used to introduce a sparsity pattern in $A$, but there is a trade-off between sparsity and accuracy~\cite[Chapter 12]{Fasshauer}, and the support radius parameter needs to be carefully adjusted. Moreover, a different choice of RBFs might interfere with the basic requirements to avoid oscillatory behavior. 
A fast multipole expansion is another powerful method which can be used for a fast evaluation of RBF interpolants but requires a large number of points to pay off~\cite{FMM2D,FMM3D,FMM4D}.
A stable and simple to implement method is the so called PU method \cite{Wendland02}, which can be understood as generalization of Shepard's interpolation method \cite{Shepard68}. 
The interpolation domain is covered by $N_P$ overlapping ($d$-dimensional) patches.
The global interpolant $s(\vc{x})$ is then given by
the weighted sum over all $N_P$ local patches.
\begin{equation}
\label{eq:PU}
s(\mathbf{x}) = \sum_{i=1}^{N_P} w_i(\mathbf{x}) u_i(\mathbf{x})\,,
\end{equation}
The local interpolant $u_i(\vc{x})$ for every patch is a PHS interpolant
(see eq.~\eqref{eq:PH}).
The PU weights $w_i(\vc{x})$ are then found by using Shepard's method.
\begin{equation}
w_i(\vc{x}) = \dfrac{\psi_i(\vc{x})}{\sum^{N_P}_{l=0} \psi_l(\vc{x})}
\end{equation}
The generating functions $\psi(\vc{x})$ used here are Wendland's $C^2$
functions \cite{Wendland95} with compact support: 
\begin{equation}
\label{eq:psi}
\psi_{i,d}(\vc{x})\equiv\psi_d\left(\dfrac{\|\vc{x}-\vc{c}_i\|}{\rho_S}\right)
=\psi(\rho_i).
\end{equation}
Here $\psi(\rho_i)$ is an RBF with support on the patch around the patch center $\vc{c}_i$
with a support radius $\rho_S$. The specific choice of $\psi_d$ \cite{Wendland95} depends
on the dimensions $d$ of the interpolant. For $d=2$ and $d=3$ $\psi_d$ is
\begin{equation}
\psi = \left(4\rho_i + 1 \right) \left( 1-\rho_i \right)_+^4\,,
\end{equation}
while for $d=4$ and $d=5$ it is
\begin{equation}
\psi = \left(5\rho_i + 1 \right) \left( 1-\rho_i \right)_+^5\,,
\end{equation}
(for more than 5 dimensions see \cite{Wendland95}).
The function $\psi_2(\rho_i)$ is shown in fig.~\ref{fig:WC2}.
The specific choice of $\psi$ ensures derivatives of order two to be continuous at the
patch boundaries where two or more patches overlap.
The notation $(\cdot)_+$ means that the function is non-zero only for a positive argument and
thereby introduces the compact support.
By choosing the support radius $\rho_S$ such that only a subset of the sample points is covered, the problem
can be handled more efficiently for a large number of sample points.
The size of the local (independent) PHS interpolation problems can now be chosen such that the linear systems of equations can be solved efficiently.
A possible
choice of patch covering is depicted in fig.~\ref{fig:part}. The specific choice
of domain centers and patch sizes will be discussed in detail in sec.~\ref{sec:alg}.

By using the combination of PHS and the PU method a hierarchical interpolation scheme is obtained which overcomes the scaling problems
of a pure PHS interpolation \cite{cavoretto2014meshless}. The splitting in localized patches also allows for an efficient parallel implementation and the extension to higher dimensions and larger number
of sample points.

\subsection{Iterative refinement scheme}
\label{sec:refinement}
The basic strategy to reduce the number of sample points needed, is to start
out with a coarse grid and to refine locally on an as-needed-basis.
The refinement procedure itself includes a scheme to generate new sample points
in way that helps to reduce the amount of necessary function evaluations.
We follow the ideas according to refs \cite{Hales02,Iske05,Her07} describing a general
scheme for the error scaling with step wise grid refinement.
The error estimates determining the refinement are done by interpolating the error.
This has the advantage that when testing for convergence, the function itself 
does not need to be evaluated.

\subsubsection{Refinement process}
In the following we describe the general procedure of the iterative refinement.
The specific choice of sample points will be described in detail in the next section.
Here the interpolation scheme from eq.~\eqref{eq:PU} is used not only to approximate
the function but also to approximate the error $e$ of the interpolant $s_c$
in the space between the sample points~\citep{Iske05}. This has the advantage that
the function itself needs not be evaluated to calculate the error in the
new points which are considered for refinement.

For that purpose the error between the interpolant $s_c$ and the function $f$
\begin{equation}
\label{eq:ex}
e^k(\vc{x}) = 
\begin{cases}
0 &\forall \vc{x} \in X^k_c\\
s_c^k (\vc{x})-f(\vc{x}) &\forall \vc{x} \in X^k_e\\
\end{cases}
\end{equation}
is defined, where $k$ denotes the iteration index.
Here the two different point sets $X_c^k$ and $X_e^k$ are used to define the error $e$.
The sample points $X_c^k$ define $s_c$ and thus the error is zero in those points
by definition. 
The error sample points $X_e^k$ are chosen from a subset of a finer grid than $X_c^k$ and are used to measure the error of $s_c^k$ with respect to $f$.
An interpolant for the error $s^k_e$ can then be constructed from $e^k$ and the points $\{X_e^k,X_c^k\}$. 

To refine further, a set of refinement points $X_r$ is chosen and the error in these points is estimated by $s^k_e$.
To decide which function values are required to create an improved interpolant $s_c^{k+1}$, satisfying
a given error bound $\epsilon$, the estimated error $\lvert s_e(X_r) \rvert$
and the calculated errors $\lvert s_c (X_e)-f(X_e) \rvert$ have to be checked
against the error threshold value $\epsilon$. All points with an error exceeding $\epsilon$ are
added to the set $X_c^{k+1}$ and eventually the next iteration can be
performed. The procedure has converged if no further points have been added to $X_c$,
or is terminated if a maximum number of iterations is reached.

The function $f$ itself is only evaluated at the points $X_c$ and $X_e$,
but not at the refinement points $X_r$.
Since the number of points grows polynomially
during the refinement process the saving in terms of function evaluations of $f$ is
substantial. The errors in the refinement points $X_r$ are estimated by interpolation and thus not
necessary to evaluate directly. Hales et al.~\cite{Hales02} predicts linear
convergence of the error with respect to the refinement level for  a regular dense grid directly with PHSs.
Even though this might not be the case for a non-regular grid as is produced
by the algorithm presented here, the locally refined areas can be expected to show a similar convergence behavior.

\subsubsection{Generation of grid points}
In the following the specific choice of the sets of sample points $X_c$, $X_e$, and $X_r$ will be introduced.
Possible schemes for choosing sample points are, e.g., low discrepancy points like Halton points \cite{Halton60} or points that are suitable for polynomial approximations like Chebyshev points.
Sparse grid methods~\cite{Gerstner08} combine results computed on a particular sequence of structured grids in order to get the final approximation. This type of technique has also been used together with RBFs~\cite{GLS13}.

There are various types of point sets that have good properties for interpolation. However, here the choice of a simple regular grid greatly simplifies the creation and implementation of a systematic refinement scheme.
The regular grid is expressed in terms of simple basis point sets
which allow for a recursive refinement and a systematic splitting into local cells. 

We use a dense grid to define the point sets, but use only those sample points that are necessary (as explained in the previous section). A simple choice is a so called product grid.
Let $G^{(\vc{\ell})}(L)$ be a $d$-dimensional grid on the hyperrectangular domain $L$ defined by its side lengths $L_i$, $i=1,2,\ldots,d$, and let $\vc{\ell}=(\ell_1,\ell_2,\ldots,\ell_d)$ be a multi-index describing the refinement.
The step width between the grid points in
a certain direction $i$ is then given by $L_i 2^{-\ell_i}$.
The total number of grid points is given by $\left(2^\ell+1\right)^d$ if $\vc{\ell}$ is equal for all $i$.
In the  local refinement scheme, a global point set of refinement level (depth) $\ell$ contains a subset of all points from the grid $G^{(\ell)}(L)$.

\begin{figure}
\includegraphics[width=4cm]{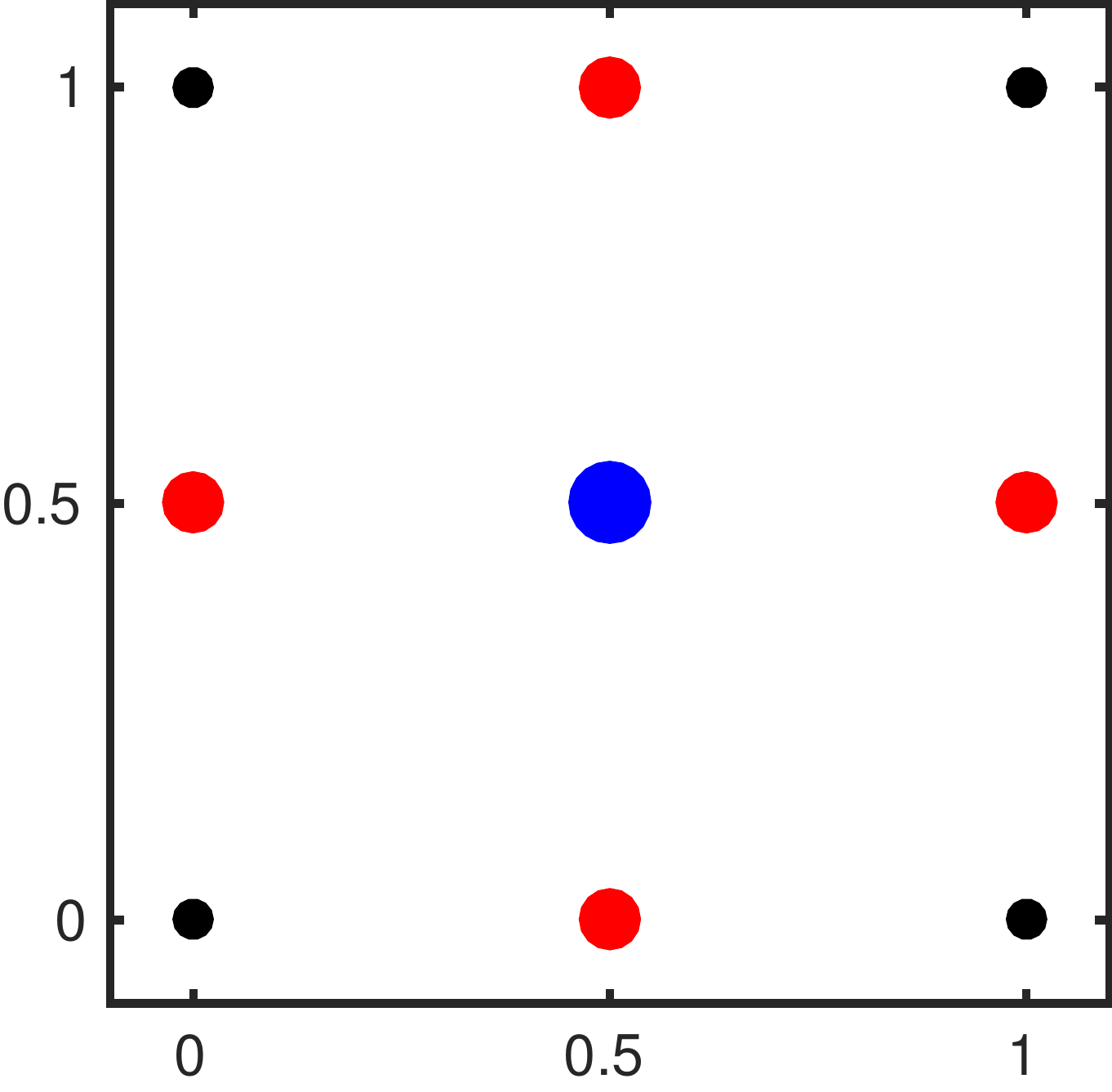}
\includegraphics[width=4cm]{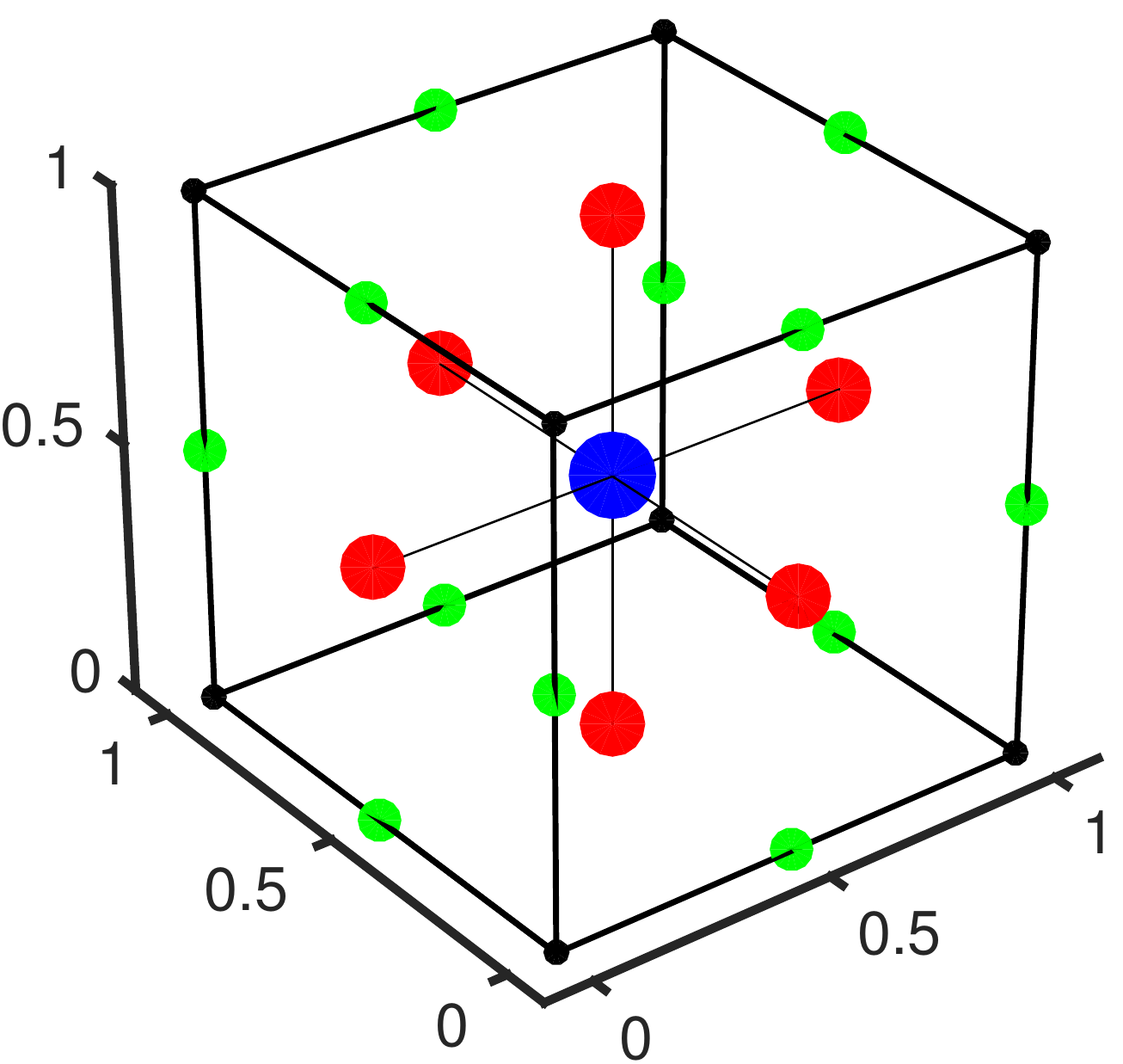}
\caption{Basis points for a product grid in a 2D (left) and a 3D cell (right).
The functions are coded by color. 2D: black $V^{(0)}$, red $V^{(1)}$, blue $V^{(2)}$.
3D: black $V^{(0)}$, green $V^{(1)}$, red $V^{(2)}$, blue $V^{(3)}$. Note that both grids represent
an $\ell=1$ refinement of a cell.}
\label{fig:pgrid}
\end{figure}

A hierarchical cell structure allows for organizing a local refinement,
and can reduce the number of sample points, especially in higher dimensions.
A cell $C$ is defined by its $2^d$ corner points, which in the following will be denoted by $V^{(0)}$, and forms the first basis point set for a cell (see fig.~\ref{fig:pgrid}).
To allow for local variations in the refinement we prefer to define the grid recursively by only using refinement levels $\ell=0$ and $\ell=1$ within a single cell. A fully refined cell, containing all the points of level $\ell=1$ can be split into $2^d$ new cells with refinement level $\ell=0$ forming the hierarchy of cells.

The local dense level $1$ grid $V\equiv G^{(1)}(C)$ in a cell, as illustrated in fig.~\ref{fig:pgrid}, is composed by $d$ different basis point sets,
$V = \bigcup_{i=0}^d V^{(i)}$.  The first point set $V^{(0)}$ represents corner points (as indicated above), $V^{(1)}$ represents edge centers, $V^{(2)}$ represents face centers, and so on. For convenience and to generalize the concept to an arbitrary number of dimensions we introduce a mathematical description for 
the basis points in a cell of unit side length. Let $r_j$, $j=1,2,\ldots,d$ be indices that can take the values 0, 1 or 2. 
\begin{equation}
b_r = \dfrac{1}{2}(r_1,\ldots,r_d).
\end{equation}
In a grid consisting of several cells, we can also define the unique points within a cell as
\begin{equation}
b_r^\prime= \dfrac{1}{2}(r_1,\ldots,r_d)\,,
\end{equation}
where the indices $r_j$ can only take the values 0 or 1.  

A set of basis points $V^{(n)}$ is defined as all permutations of $b_r$ with $n$ odd indices ($n$ indices $r_j=1$). This can be understood in terms of shifts. $V^{(0)}$ contains the unshifted corner points. One shift of half a cell length in any dimension gives a point in $V^{(1)}$, which is then an edge center. The single point in $V^{(d)}$ after $d$ shifts is always the cell center. The number of points in $V^{(n)}$ in a single cell is given by $\binom{d}{n} 2^{d-n}$.
The basis point sets $V^{(n)}$ in two and three dimensions are  illustrated in fig.~\ref{fig:pgrid} by differently colored dots. 

\begin{figure}
\includegraphics[width=8cm]{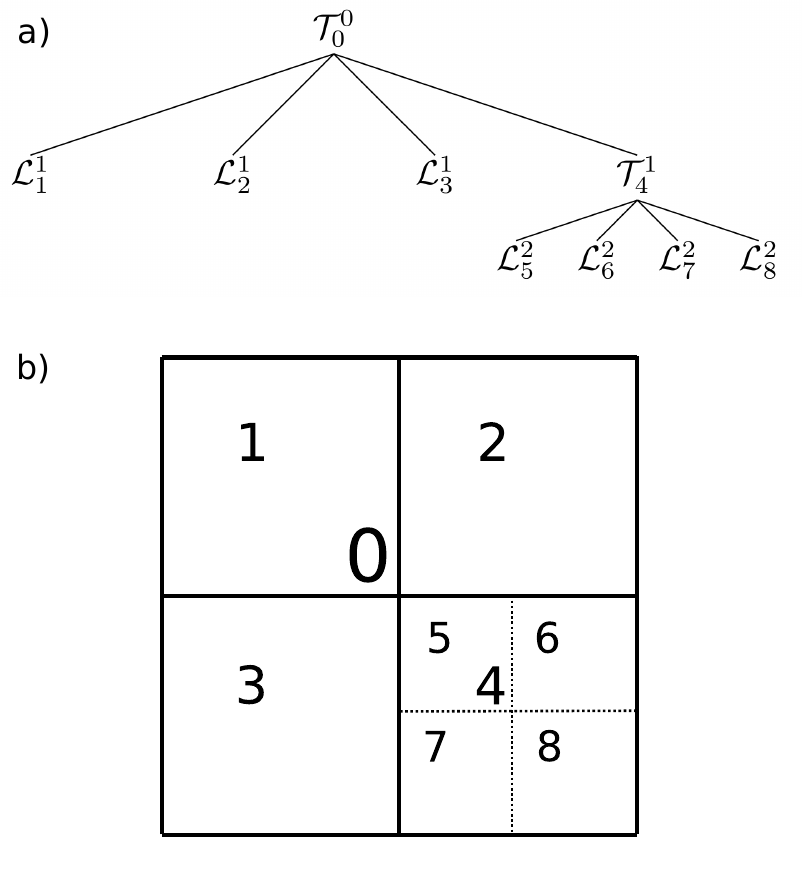}
\caption{Sketch of the cell structure and their subdivision. In (a) a tree
with a local refinement depth of $\ell=2$ is shown. The corresponding grid is
depicted in (b). The cells are linearly indexed: subcripts in (a) and numbers in (b).}
\label{fig:tree}
\end{figure}

To describe a locally refined grid a tree structure is used.
The tree ${\cal T}^{(\ell)}$ of depth $\ell$ contains a subset of all points from the grid $G^{(\ell)}(L)$: ${\cal T}^\ell \subseteq G^{(\ell)}(L)$. The tree itself is built from its nodes ${\cal T}_i^j$, $j=0,1,\ldots,\ell$, where ${\cal T}_0^0$ is the root node, and the range of $i$ depends on the number of cells at each level. Leaf nodes are denoted by $\leaf_i^j$. Each leaf node holds the information about its points. For an unrefined leaf cell, only the $V^{(0)}$ point set is present. Under refinement, further basis point sets are added until the correpsonding cell is fully refined, at which point the cell is split and another level of leaf nodes are created with the previous node as parent.
The mapping between the nodes, leafs and
the corresponding cells is sketched in fig.~\ref{fig:tree}.
A patch in the interpolation scheme covers a subtree with an appropriate number of associated points at its leaf nodes, see fig.~\ref{fig:part}.

For the refinement process the choice of sample and check points can now be derived
from the point generation scheme described above. The set $X_c$ defining the interpolant is initially formed by the coarse grid $G^{(\ell_0)}(L)$. The error check points $X_e$ are then all points in $V^d$ of all leafs nodes,
namely the center points of all cells. The error estimate points $X_r$ are given by $V^{(d-1)}$. In case of $d>2$ the next step would be $X_e \in V^{(d-1)}$ and
$X_r \in V^{(d-2)}$ and so on as long as $d-j>0$.
A fully refined cell can then be split into $2^d$ new cells and the recursion
depth increases. Note that the points $X_c$ used in the interpolant do not need to be all
the points contained in ${\cal T}$, rather a subset creating a sufficient
approximation $s_c$ of $f$.

\subsubsection{Algorithm}
\label{sec:alg}
We construct an algorithm which
considers the special features of the PHS--PU scheme and incorporates the refinement
process. We also address some practical issues in the algorithm.
A formal outline of the algorithm is given in alg.~\ref{alg:PSLR}.
The main goal of the algorithm is to reduce the number of function evaluations of the
function $f$ and at the same time provide a method which gives an
reliable error estimate. In the final application $f$ will be a quantum
chemistry program and $\vc{x}$ represents a molecular geometry in internal
coordinates. Calculating
$f(\vc{x})$ means to carry out a single point energy calculation, which is
considered as the computationally expensive part.
In essence the algorithm is designed to locally refine the grid as needed to represent the PES with a desired accuracy. Moreover,  most applications
only require a certain energy range of interest. This can be taken into account in the refinement scheme and provides a further improvement. 

To start the refinement process a function $f(\vc{x})$ and initial set of points $X_c$ has to be chosen. It is assumed that a regular grid
with a low refinement level $\ell_0$ is used. The level $\ell_0$ depends on the chosen
order of the PHS and the size of the domain or the function $f$.
The level $\ell_0$ should be chosen such that it has enough points for the
linear system eq.~\eqref{eq:sys} to be well posed and such that the qualitative features of $f$ are not
undersampled. For most application $\ell_0=2$ (5 points in each directions) might be a good starting value.

The main refinement loop consist of two nested loops -- the outer loop
refining $\ell$, while the inner loop refines the point set $V$ within the cells.
In the inner loop the error function $e(\vc{x})$ from eq.~\eqref{eq:ex}
is created from the sample points $X_c$ and the newly generated error check points $X_e$. For all points in $X_e$ the function $f$ has to be invoked
in order to evaluate the interpolated error (eq.~\eqref{eq:ex}).
Error check points which are above the threshold join the set $X_c$ for the next iteration and contribute to a refined grid. Error checkpoints which are converged
can remain in $X_e$, as this may improve the quality of the error estimator.
The next step is to evaluate the approximate error $s_e(\vc{x})$ at the refinement points $X_r$ and
add the points above threshold to $X_e$ for the next round. Note that to identify those points the function $f$ is not needed. This happens in the next step for the
points in $X_r$ which are predicted to be not converged.
After the inner loop is done non-converged $X_r$ points are
added to $X_c$. If there were no new $X_c$ points produced the
refinement process can be considered as converged.

For example for a 3D system the initial choice of points in a single in the inner loop would go as follows. The point set $X_c$ consists of the all corner points of the cube ($V^{(0)}$) and the
error check points are the center points of the cube ($V^{(3)}$). $V^{(d)}$ contains only a single point for each cell and has the largest possible distance from the corner points. The error estimate is then calculated for all points on the faces of the cube ($X_r = V^{(2)}$).
The error estimation $X_e$ is then improved with all non-converged points from $X_r$
In the next and last round the error is estimated for all points sitting on the edges on of the cube  ($X_r = V^{(1)}$).

After the iterations over the inner loop are finished,
all active cells $\leaf$ which are subject to refinement are inspected.
If a cell did not produce any new points it can be considered as locally converged
and can be taken out of the refinement process. If all points of a cell are above
a given energy threshold it can be also declared converged, since it is not
in the region of interest. Also if all points in a cell produced invalid
function values $f$ there is no need to refine them any further.
This might however indicate severe problems of the electronic structure calculation in this region of space.

Now all the remaining leaf cells $\leaf_r$ are split up to produce a set
of child cells which contain the newly generated points of the last round.
Since the number of points in $X_c$ used to build $s_c$ has increased it has
to be checked that the number of points in the patches do not exceed a given
limit. Otherwise the patches have to be subdivided accordingly. To keep
the method simple a patch is built from a cell in the tree. Only
the number of points in the cell (and its child cells) is counted to check the limits rather than the number of points in the final hypersphere enclosing the cell. The point limit
$N_{p,lim}$ has to be such that it contains enough points to ensure a well posed
local PHS interpolant after splitting the patch. 

After the procedure has reached convergence the result is a set of points $X_c$
which defines the interpolant $s_c$. There are still points
in $X_e$ for which $f$ has been evaluated. These points can in principle
be added to $X_c$ to improve the quality of $s_c$ at no further cost.
We denote this merged point set by $X_c^f$.
In the results we will also show a comparison of both variants.

\begin{figure}
\begin{algorithm}[H]
\caption{The refinement algorithm}
\begin{algorithmic}
\Require Initial set of points $X_c^0 \in G	^{(\ell_0)}(L)$, $\epsilon$, $f$.
\State Create an initial set of error checkpoints $X_e^0 \in V^{(d)}(\leaf)$.
\State $k=0$
  \While{$\ell < \ell_{limit}$}
\State $X_e^k \leftarrow \{X_e^k \cup V^{(d)}\} $
\For{$n=d-1~\text{to}~1$}
\State construct $s_c$ from $X_c^k$.
\State evaluate errors $e(\vc{x})$ at $\{X_e^k \cup V^{(n+1)}\} $ (see eq. \ref{eq:ex}).
\State construct $s_e$ from $\{X_c,X_e\}$.
\State $X_c^{k+1}\leftarrow \{X_c^k,\vc{x}\}\quad \forall \vc{x} \in X_e^k \wedge \lvert s_c (\vc{x})-f(\vc{x}) \rvert > \epsilon$
\State $X_e^{k+1}\leftarrow \{\vc{x}\}\quad \forall \vc{x} \in X_e^k\wedge \lvert s_c (\vc{x})-f(\vc{x}) \rvert \leq \epsilon$
\State Create error control points: $X_r \leftarrow V^{n}$.
\State $X_e^{k+1}\leftarrow \{X_e^{k+1},\vc{x}\}\quad \forall \vc{x} \in X_r \wedge \lvert e(\vc{x}) \rvert > \epsilon$
\EndFor
\State $X_e^{k+1}\leftarrow \{\vc{x}\}\quad \forall \vc{x} \in X_r \wedge \lvert e(\vc{x}) \rvert < \epsilon \cup X_e^{k+1}$
\State $X_c^{k+1}\leftarrow \{\vc{x}\}\quad \forall \vc{x} \in X_r \wedge \lvert e(\vc{x}) \rvert > \epsilon$ \
\If{$\{X_c^{k+1} - X_c^{k}\}= \emptyset $ }\State quit\EndIf
\State Remove all converged leaf cells from refinement process.
\State Split remaining leaf cells, and
\State create new cells with $2^d$ corner points.
\State Split all patches with $N_p > N_{p,limit}$.
\State k $\leftarrow$ k+1
\EndWhile
\end{algorithmic}
\label{alg:PSLR}
\end{algorithm}
\end{figure}

\section{Results}
\label{sec:results}
To demonstrate the performance of the method a convergence study is conducted
on a model function, which resembles a situation commonly found in molecules. Moreover an example for a PES produced by an electronic structure calculation will presented.
\subsection{Model Potentials}
The workhorse for evaluating the method will be a Morse-potential:
\begin{equation}
\label{eq:fx}
 f(\vc{x}) = \dfrac{1}{d}\sum_{i=1}^d \dfrac{g_i(x_i)}{g_i(0.2)}
 \end{equation}
 with
 \begin{equation}
 g_i(x_i) = \left(1 - \exp(-(1+0.1i)(x_i - 1)) \right)^2
  \end{equation}
The function is chosen to be normalized between 0 and 1 in the investigated range ($x_i \in [0.2;5]$). Moreover the
anharmonicities are chosen to be different for every dimension to break up symmetry.
The Morse potential offers a simple but realistic scenario, to challenge the
performance of the method.
It provides high curvatures as well as flat areas which is useful to test the
local refinement properties of the algorithm.

The interpolation errors are estimated by comparing at a set $X_t$ of 3000 random generated points versus the real function value. The figure of merit giving a measure
for the reliability of the error estimates is the maximum error:
\begin{equation}
\label{eq:Emax}
 \| E\|_\infty  = \max_{\vc{x} \in X_t} \lvert f(\vc{x}) - s_c(\vc{x}) \rvert\,.
\end{equation}
A measure for the global quality is the average error:
\begin{equation}
E_{avg} = \dfrac{1}{N}\sum_{\vc{x} \in X_t} \lvert f(\vc{x}) - s_c(\vc{x}) \rvert
\end{equation}

\subsubsection{General convergence of the PHS--PU scheme}
In a first test, the convergence with respect to refinement is tested. No
local refinement is used, but dense grids with different $\vc{\ell}$ are evaluated.
The errors here are not evaluated at random points but at the newly introduced points from the next refinement level $\ell+1$.
\begin{figure}
\includegraphics[width=8.5cm]{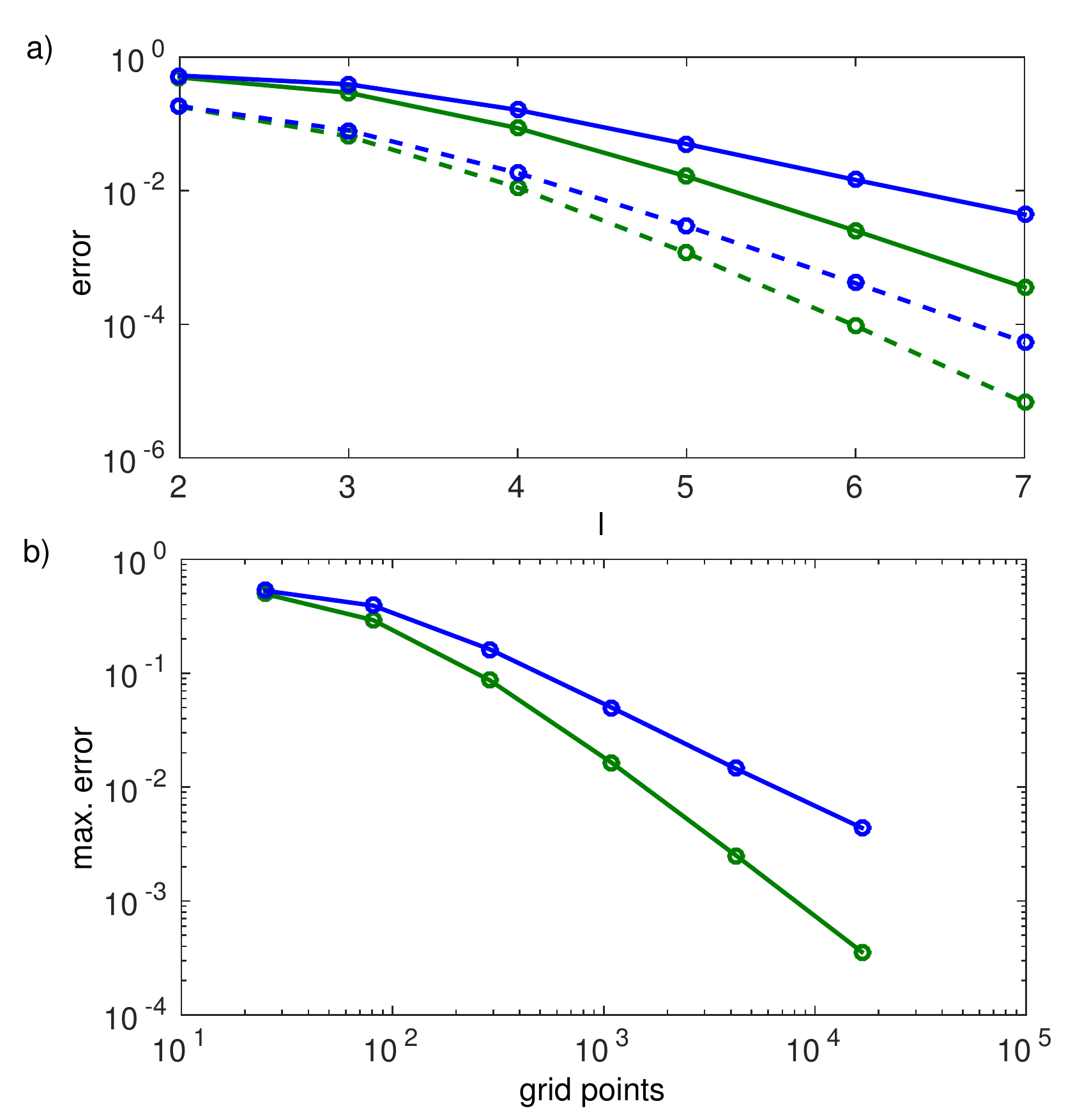}
\caption{Comparison of PHSs with order $m=2$ (blue) and $m=3$ (green) in two dimensions ($d=2$) without adaptive refinement. In (a) the maximum error (solid line) and average error (dashed line) versus the grid refinement
level $\ell$ is shown. In (b) the number of points vs.\ maximum error is shown.}
\label{fig:full_err_conv}
\end{figure}
In fig.~\ref{fig:full_err_conv} the result is shown for the  2D-Morse function
from eq.~\eqref{eq:fx}. Here we also compare second (blue) and third order (green) PHS. In fig.~\ref{fig:full_err_conv}(a) the maximum (solid lines) and mean errors (dashed lines) are shown.
Comparing the maximum error and the average error shows that independent of the
order, the rate of convergence for the average error is higher, and at $\ell=7$ the average error is about
one order of magnitude better than the maximum error.
It can be seen that the convergence rate for the third order PHS is significantly better than for $m=2$. The use of a $m=3$ PHS is thus preferred, since the computational cost is only slightly higher,
but the result much better. At $\ell=7$ the maximum error is one order
of magnitude better. In fig.~\ref{fig:full_err_conv}(b) the total number of
grid points is plotted against the maximum error. This gives an impression of the
increase of efficiency with increasing spline order.
In principle even higher orders might be used but with increasing length of the
polynomial part ${\cal P}^m$ the minimum number of points needed to start the refinement algorithm is also increased. However, this becomes more challenging with an
increasing number of dimensions.

\subsubsection{Demonstration of the local refinement process}

\begin{figure*}
\includegraphics[width=17cm]{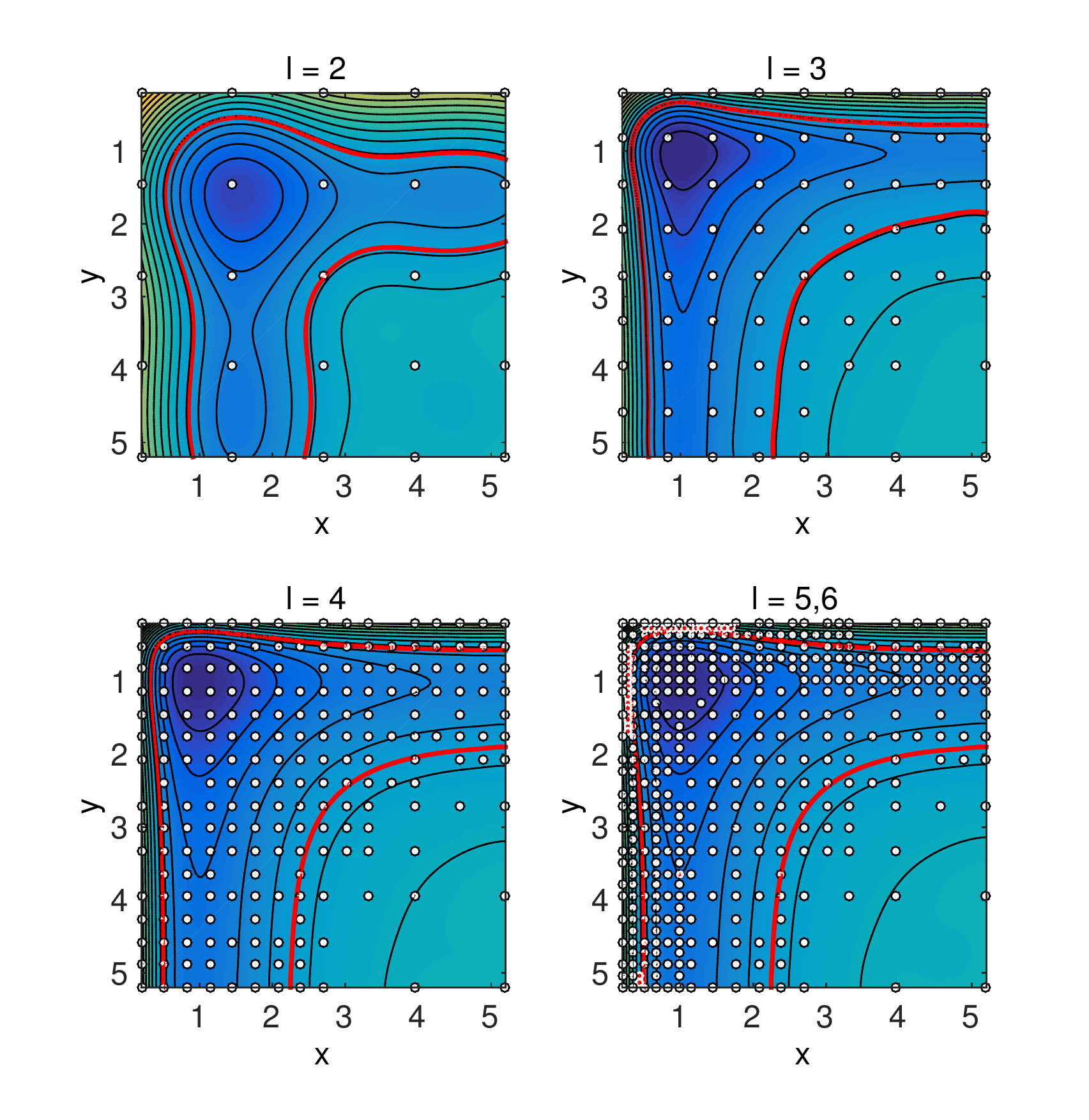}
\caption{Visualization of the iterations in the refinement process for
different refinement levels $\ell$. 
The sample points of the interpolant are indicated by the white points.
The red line indicates the contour of $f_{cut}=0.33$. The additional points introduced at the
refinement level $\ell=6$ are indicated by the white, red filled points.
The requested error is $\epsilon=10^{-3}$.}
\label{fig:Morse_Movie}
\end{figure*}

In fig.~\ref{fig:Morse_Movie} the steps of the iterative refinement (alg. \ref{alg:PSLR}) are visualized for the
Morse potential in two dimensions over all refinement steps until convergence is
reached.
We also make use of the feature to exclude regions of high energy on the PES to save function evaluations. The energy cut-off of $f_{cut}=0.33$ is indicated by the red
line. The desired accuracy is set to $\epsilon=10^{-3}$ and 
the order of the PHS used is set $m=3$ from here on.
Already in the $5\times5$ ($\ell=2$) grid of sample points, the basic qualitative features of the
PES become visible.
At $\ell\geq3$ benefits of energy thresholds become visible. The flat areas in the
lower right corner of the coordinate system are not refined any further.
However they are described with sufficient accuracy and do not alter the qualitative
shape of the PES.
In the steps with $\ell>4$ the refinement is only needed at the outer regions
close to the potential wall where the curvature of the function is high.
It is noteworthy that the refinement points chosen here are mostly points on the edge of the cells and not their midpoints. This results in a local structure which
is similar to a sparse grid \cite{Gerstner08}.
In the last step $\ell=6$ only a few new points are generated. They are located
at the potential wall close to the minimum and support a steep region of high curvature and high curvature change rates.
The result is a surface which is described by 428 sample points (the number of points in $X_c$). The total of amount function evaluations in this case was 799
(the number of points in $X_c^f$).

\subsubsection{Convergence with local refinement}
First we look at the reliability of the algorithm with respect to the
requested error $\epsilon$ or the convergence criteria. The choice of grid
points for error checking is an approximation which includes the
assumption that the errors are largest in those points.
For the following numerical test we make full use of the local refinement procedure
and the PHS--PU scheme. The maximum number of points before a patch is split has
been set to $N_{p,lim}=300$.

\begin{figure}[t]
\includegraphics[width=8.5cm]{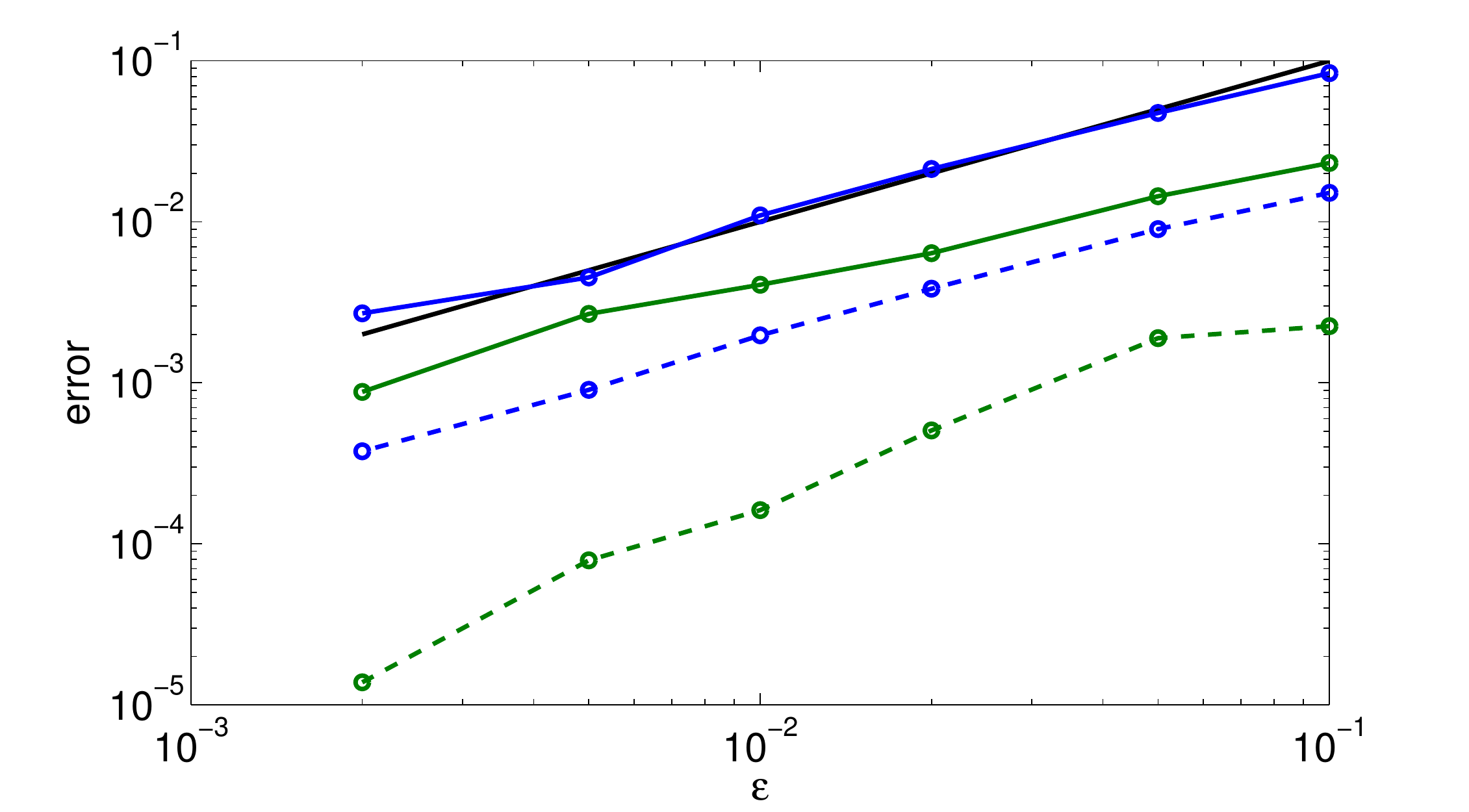}
\caption{Error measures vs.\ the requested error. Shown are the maximum errors (solid lines) and average errors (dashed lines) for a 3D Morse function. The two point set versions $X_c$ (blue) and $X_c^f$ (green) are shown. The black line indicates the requested limit $\epsilon$.}
\label{fig:Morse_3D_errors}
\end{figure}
In fig.~\ref{fig:Morse_3D_errors} the maximum error
$\|E\|_\infty$ (solid lines) and the mean error $E_{avg}$ (dashed lines) are plotted versus the requested error $\epsilon$ for the point sets $X_c$ and $X_c^f$.
In the case of the maximum error it can be seen that the error can be larger
than $\epsilon$, but is close to $\epsilon$ in the tested examples. 
However, $\|E\|_\infty$
scales accordingly with $\epsilon$ and gives a similar error.
In contrast the average error is significantly better and well below the given
error bound.
The error study for a full grid refinement as shown in fig.~\ref{fig:full_err_conv}
already indicates that due to the general convergence of PHSs the next
refinement level yields a major improvement in the error.
The dashed lines in fig.~\ref{fig:Morse_3D_errors} represent the results when the
calculated error control points are incorporated into the sample points ($X_c^f$).
In this example the maximum error is lowered by one order of magnitude compared with when
using $X_c$. The average error is also improved significantly and is almost two orders of magnitude better than the requested error $\epsilon$.
Its thus highly beneficial to use the full point set $X_c^f$. This means
that the errors are significantly lower than $\epsilon$.
In the following the $X_c^f$ point sets are used in the results if not
 otherwise stated.

\begin{figure}[t]
\includegraphics[width=8.5cm]{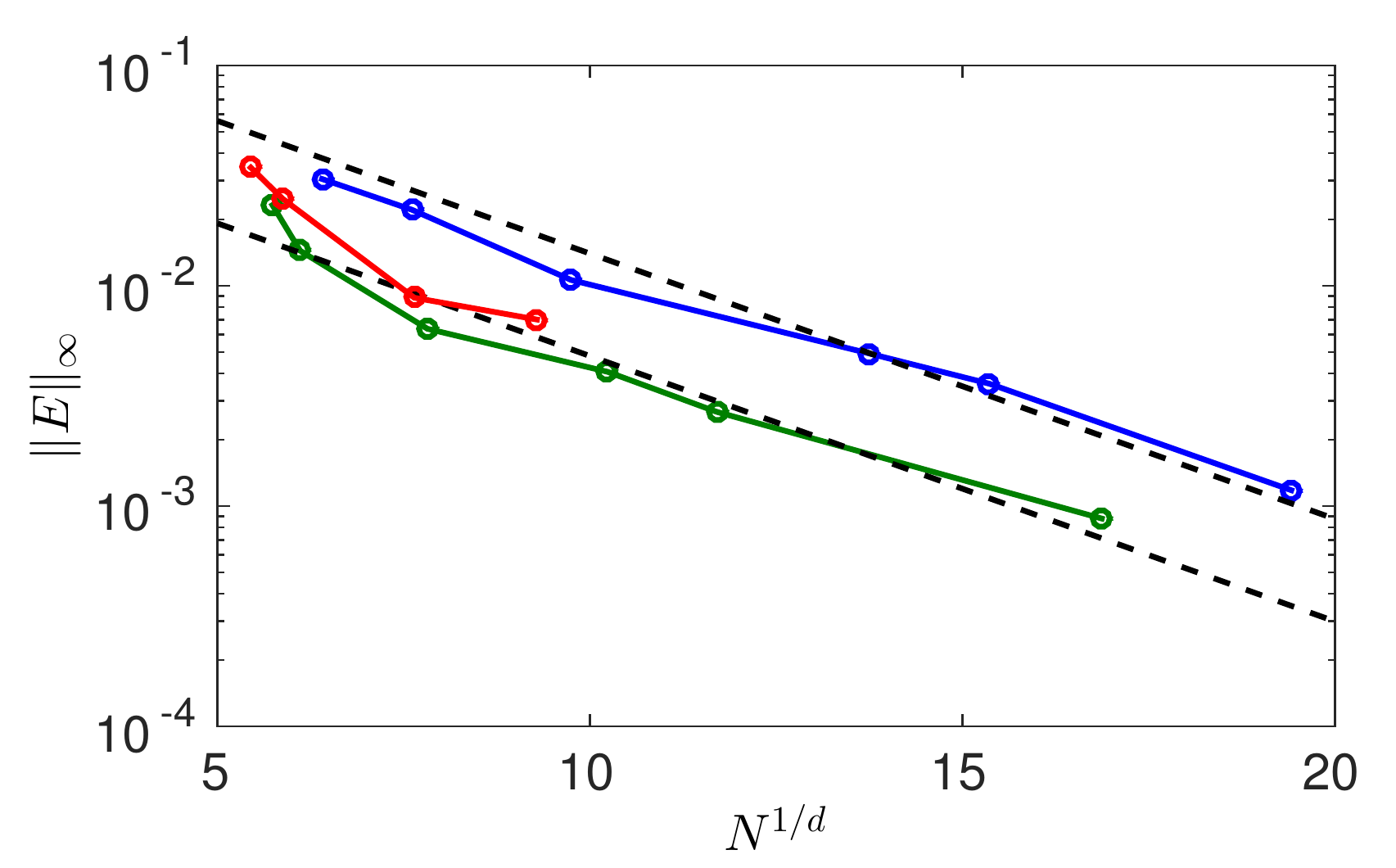}
\caption{Maxmimum error vs.\ the number of total points ($X_c^f$) per direction. Comparison of 2D (blue), 3D (green), and 4D (red). The dashed line indicates third order convergence.}
\label{fig:Morse_ND_pts}
\end{figure}
Another interesting question is the scaling behavior with respect to the
number of dimensions. The underlying interpolation method and the refinement
process are designed to work for any number of dimensions ($d \geq 2$).
In the following we interpolate on a smaller domain such that $x_i\in [0.3;3.3]$
and compare the scaling for $d=2,3,4$. In fig.~\ref{fig:Morse_ND_pts}
the maximum error is plotted against the average number of points per dimension $N^{1/d}$ which is proportional to the average fill distance of points~\cite{Fasshauer}.
The dashed lines in fig.~\ref{fig:Morse_ND_pts} indicate the expected third order convergence~\cite[p.~129]{Fasshauer}. The improvement of the maximum error is defined by the following equation:
\begin{equation}
\label{eq:conv}
\lim_{n\rightarrow\infty}\|E_{n+1}\|_\infty =  C \dfrac{\|E_{n}\|_\infty}{N_{n}^{r/d}}\,,
\end{equation}
where $r$ is the rate of convergence. It can be seen that for all dimensions shown third order convergence can be achieved. For 3 and 4 dimensions the necessary
number of points per direction is even decreased. 
This result provides valuable information and can be used to extrapolate to
the necessary number of points needed for a certain accuracy by using
a coarse interpolant.

\subsubsection{Efficiency of the local refinement}
The iterative refinement process is designed to use localized error estimates
to decide where refinement of the sampling grid is needed. By only refining on
an as-needed-basis the method is expected to be more efficient than an overall
refinement on a dense grid.
The efficiency of the local refinement is evaluated by the maximum refined
level $\ell_{max}$ a calculation has reached. We then define the efficiency as
the ratio of the actual number of sample points needed per dimension $N^{1/d}$ and the number of points per dimension generated in a full dense grid of level $\ell_{max}$:
\begin{equation}
R = \dfrac{N^{1/d}}{(2^{\ell_{max}}+1)}\,.
\label{eq:eff}
\end{equation}
\begin{figure}[t]
\includegraphics[width=8.5cm]{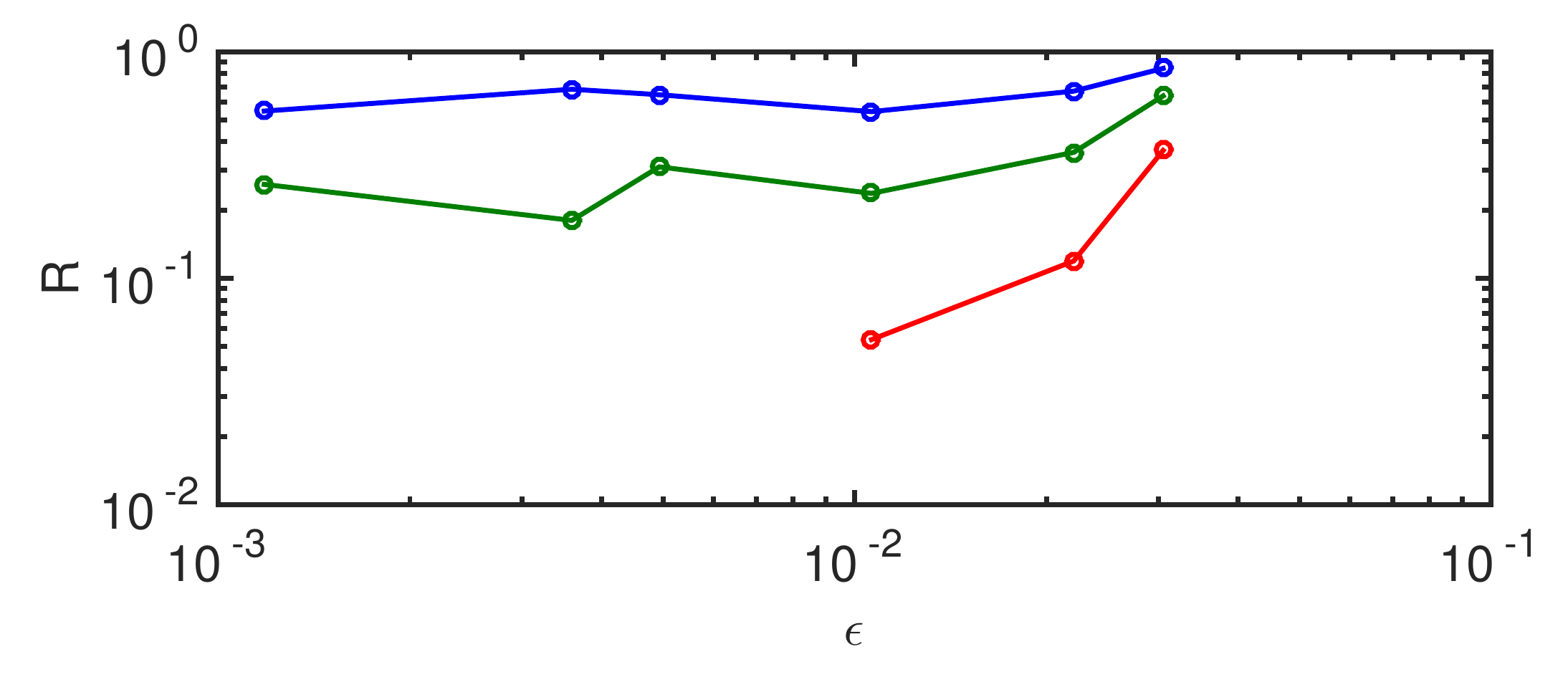}
\caption{Efficiency of the refinement method vs.\ the maximum error. Comparison of $d=2$ (blue), $d=3$ (green), and $d=4$ (red). Smaller numbers are better.}
\label{fig:Morse_ND_eff}
\end{figure}
Figure \ref{fig:Morse_ND_eff} shows the efficiency as a function of the
maximum error for $d=2, 3, 4$. With an increasing number of dimensions the
saving in terms of sample points to represent the PES increases.

\subsection{Molecular PES from ab intio data}
To demonstrate the interpolation method on a more realistic 3-dimensional example,
we test it against an example as it might used in
a reactive scattering calculation \cite{Kowalewski14JPCA}.
Here the PES for the backside attack of the nucleophilic substitution reaction 
$\mathrm{Cl}^- + \mathrm{CH_3F}\rightarrow \mathrm{ClCH_3} + \mathrm{F}^-$ is calculated.
The single point energies are  calculated at the MP2/6-311+G* level of theory with the ab initio program
package Gaussian09 \cite{g09}.
The active coordinates are the carbon-chlorine distance $r_{C-Cl}$, the
carbon-fluorine distance $r_{C-F}$, and the umbrella angle $\theta$ of the
$\mathrm{CH_3}$ group, where $\theta=90^\circ$ corresponds to a
planar $\mathrm{CH_3}$ group.
The system is assumed to preserve C$_\mathrm{2v}$ symmetry
during the course of the reaction and thus Cl, C, and F are in a linear configuration.
The range over which the PES is calculated is given in tab.~\ref{tab:sn2vars}
\begin{table}
\caption{Ranges for the calculation of the PES\label{tab:sn2vars}}
\begin{tabular}{lrr}
\hline\hline
variable & min. & max. \\
\hline
$r_{C-Cl}$ & 1.1\,$\text{\AA}$ & 5.1\,$\text{\AA}$ \\
$r_{C-F}$  & 0.8\,$\text{\AA}$ & 4.8\,$\text{\AA}$ \\
$\theta$   & 40$^\circ$  & 100$^\circ$ \\
\hline\hline
\end{tabular}
\end{table}

The iterative refinement process according to alg.~\ref{alg:PSLR} for the PES is performed
to obtain a set of sample points and their corresponding values defining the 3D-PES interpolant.
Each function evaluation $f(r_{C-Cl},r_{C-F},\theta)$ calls the ab initio program with the
corresponding molecular geometry and returns the corresponding energy value.
The only tunable input parameters used here are the relative error tolerance ($\epsilon = 0.01$ Hartree), and the cutoff value ($0.2$ Hartree, with respect to minimum), which are chosen to obtain a fast convergence of the procedure and only focus on the relevant energy range.
Running the automated interpolation procedure results in a set of 5008 points (the $X_c^f$ point set).
\begin{figure*}
\includegraphics[width=1.\textwidth]{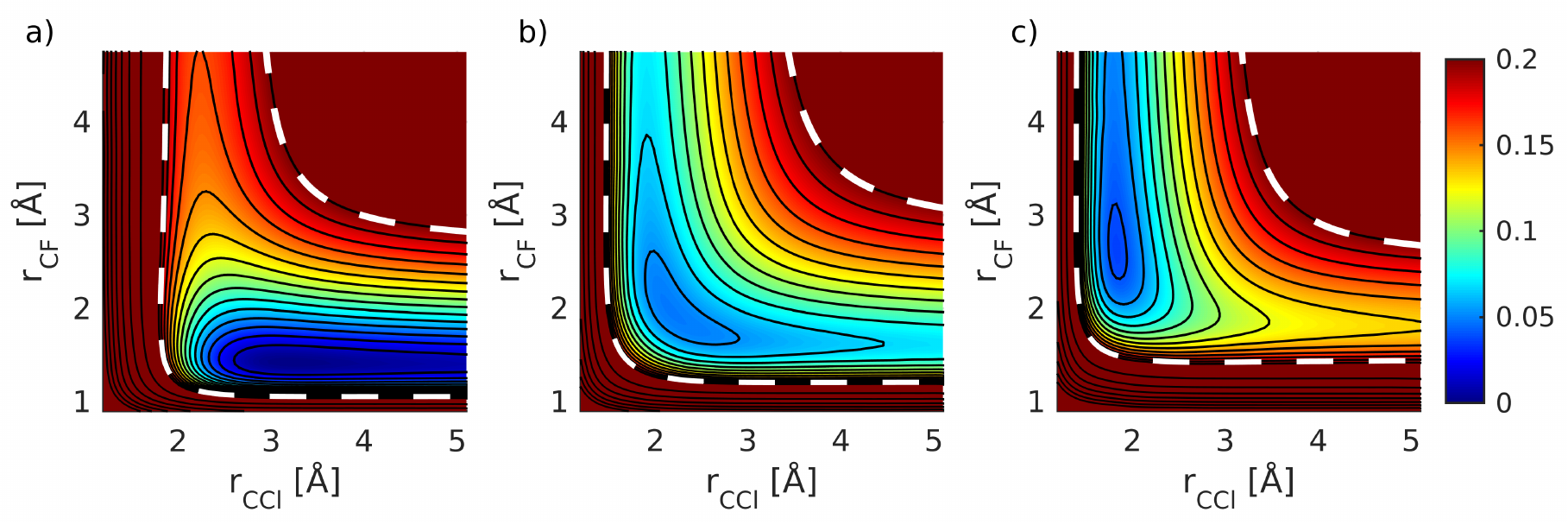}
\caption{2D slices through the interpolated PES of the nucleophilic substitution reaction $\mathrm{Cl}^- + \mathrm{CH_3F}\rightarrow \mathrm{ClCH_3} + \mathrm{F}^-$ for different values of $\theta$: (a) $71.5^\circ$, (b) $95^\circ$, and (c) $108^\circ$. The color scale is truncated above 0.2 Hartree (5.44\,eV) (indicated by the white dashed line). The spacing between the black, solid contour lines is 0.11 Hartree for energie values $>0.2$ and 0.013 Hartree for energie values $<0.2$ Hartree.}
\label{fig:mol_slices}
\end{figure*}
The interpolated PES is shown in Fig. \ref{fig:mol_slices} in form of 2D slices
for umbrella angles corresponding to the reactants, the transition state and the
products.

To evaluate the errors the interpolant is tested against a set of  4159 randomly generated ab initio
points, which  are below cutoff in the relative energy region.
The maximum error found is $\|E\|_{\infty}=1.75\cdot 10^{-3}$\,Hartree which is on the order of magnitude below the requested error bound $\epsilon$, which confirms the trend found in fig.~\ref{fig:Morse_3D_errors}.
The average error found here is $E_{avg}=7.51\cdot 10^{-5}$\,Hartree, meaning that the average
discrepancy between the interpolant and the ab initio calculation in this example is only 2.04\,meV.
The maximum refinement depth that has been reached for some cells during
the refinement process was $\ell=6$. A full, dense grid with $\ell=6$ requires
65 sample points for each dimension resulting in a total of 274\,625 sample points. The local refinement
procedure thus provides a interpolant with only 1.8\,\% of the sample points when compared to a
dense grid.

\section{Conclusion}
In this paper we have presented an adaptive interpolation algorithm for
the efficient creation of approximate global PESs.
The combination of a PHS scheme with a PU scheme allows to efficiently handle
a larger number of sample points. Ill-conditioning of the interpolation matrix is avoided.
We have demonstrated the error scaling for a the adaptive, local refinement algorithm.
The error estimation and local refinement method presented within this approach
have been tested with data sets of 2--4 dimensions.
The empirical study of the errors confirms that the error bounds are reliable.
It has been found that the average observed errors are lower by approximately
one order of magnitude than the requested error.
The local refinement makes the scheme highly efficient and places sample
points on an as needed basis. The additional energy cutoff criteria makes the
algorithm even more efficient. The number of required sample points which needed to be evaluated in the chosen examples are 
between 1-10\,\% compared to a grid of sampling points of predefined density.
Using this method in combination with state of the art ab intio methods allows for an efficient calculation of high quality global PESs.

Even though a direct calculation of higher dimensional ($>6$ dimension) seems
not be feasible with the present method, the PES for higher dimensional
problems can be approximated by a many-body expansion \cite{Quack,MOLPRO_brief,Rauhut04jcp,Hrenar07jcp}
as it is commonly used with e.g., vibrational configuration interaction methods \cite{Neff09jcp}
\begin{align}
V(q_1, \dots,q_N) = &\sum_i V_i(q_i) +  \sum_{i<j} V_{ij}(q_i,q_j)\nonumber\\
+&\sum_{i<j<k} V_{ijk}(q_i,q_j,q_k) + \dots
\end{align}
where $V_i$, $V_{ij}$, $V_{ijk}$, etc.  are the 1-mode, 2-mode, 3-mode contributions respectively. These individual components are in itself 1,2,3, ... dimensional potential energy surfaces, which can be generated in a efficient way with the presented method (Eqs. \ref{eq:PH}, \ref{eq:PU},  and  Alg. \ref{alg:PSLR}).

\begin{acknowledgements}
M.K. acknowledges support through the Centre of Interdisciplinary Mathematics (CIM), Uppsala University.
The computations were performed on resources provided by SNIC through Uppsala Multidiciplinary center for Advanced Computational Science (UPPMAX) under Project snic2013/1-267.
The research of A.H. was partially supported by the National Science Foundation (grant  DMS-1318427).
\end{acknowledgements}

\bibliographystyle{apsrev} 
\bibliography{Interpolator}

\end{document}